\documentclass[twocolumn]{article}
\usepackage{authblk}
\usepackage{graphicx}
\usepackage{xcolor}
\usepackage{hyperref}
\usepackage{natbib}
\usepackage{float}
\usepackage[margin=2cm]{geometry}

\title{\textbf{WST spectroscopic variability alerts: discovery
space, data flow system requirements\footnote{Write up of a
contributed talk, presented at SS7a {\it WST: the Wide-field
spectroscopic facility} at The European Astronomical Society
Annual General Meeting, 23-27 June 2025 in University College
Cork, Ireland.}}}

\author[1]{Valenitn D. Ivanov}
\affil[1]{\small European Southern Observatory, Karl-Schwarzschild-Str. 2,
85748 Garching bei M\"unchen, Germany; vivanov@eso.org}

\date{June 23, 2025}

\begin{document}

\maketitle

\abstract{
The enormous multiplexity of the WST opens up the possibility to
trigger alerts for variable objects -- an option that has been
reserved so far only for imaging surveys. WST can go further by
detecting spectroscopic line profile and line strength variations.
I review previous alert-issuing surveys that are limited to imaging,
and describe some of the new research possibilities that this
feature of the data flow system (DFS) would open up. The latter
range from variability of emission line stars, such as Bes, WRs
and LBVs to variability of active galaxies and quasars, including
the so-called changing look objects that shift between Type 1 and
Type 2. Furthermore, I describe the requirements that the WST DFS
must meet to make this feasible. The most critical aspect
is the rapid data processing for timely follow-up. Next, the alert
system is tightly connected with the data reduction and archive,
because it will need an extensive and continuously updated spectral
reference database. The new spectra will have to be compared against
these reference spectra to identify variations. The reference
spectra can either be ``native'' from the WST itself or they can
originate from other spectroscopic surveys. Two options for the
DFS are considered: one is to conduct an automated search of
the WST’s own archive, and potentially of other spectroscopic
archives and a second option is to allow the users to submit
reference spectra on their own. The spectroscopic alert system will
open up a completely new discovery space that is not accessible to
the existing or planned near-future surveys.
Finally, I discuss the advantages of moving the variability detection
to physical parameters by modeling the observed and reference spectra
and comparing the derived fitting parameters. This strategy offers a
robust method for alert ranking.}

\section{Introduction: Astronomy in the Next Decades}

The astronomical landscape in the next decade promises a rich
variety of facilities, both ground- and space-based, aimed at
apply many techniques, and addressing a multitude of questions.
The increasing role of wide field surveys which has became
possible a few decades ago thanks to a new generation of large
and increasingly more accessible detectors culminates in the
Rubin Observatory
\citep[started operations in 2025;][]{2019ApJ...873..111I}.
Its 16 megapixel camera covers 3.5-degree field of view and it
will deliver the closest to a continuous video-stream from the
sky that astronomy has ever approached. Right now this
instrument is the pinnacle of the time-domain astronomy which
can be traced to a number of synoptic multi-epoch surveys,
such as
MACHO \citep[MAssive Compact Halo Objects,
1992-1999;][]{1997ApJ...486..697A}
VVV \citep[Variables in the Vi\'a Lacte\'a,
2010-2024;][]{2010NewA...15..433M,2024A&A...689A.148S},
ZTF \citep[The Zwicky Transient Facility,
2018-;][]{2019PASP..131g8001G,2019PASP..131a8002B}, and
Gaia \citep{2016A&A...595A...1G}, to name a few optical and
infrared ground- and space-based milestone time-domain
projects, respectively.

Astronomers quickly realized that nearly real-time community-wide
messages about significantly varying objects, also called
``alerts'', are the key to taking full advantage of these resources.
Probably, the most successful mission to identify such variables
was Gaia \citep{2021A&A...652A..76H}. The wealth of detections
revealed a further need -- of systems called ``brokers'', able to
handle the avalanche of alerts: to classify them and set priorities.
An example is ALeRCE \citep[The Automatic Learning for the Rapid
Classification of
Events][]{2021AJ....161..141S,2021AJ....161..242F,2021AJ....162..231C}
which is currently processing the ZTF alert stream, but is
intended to handle the Rubin data.

The ecosystem of tools developed for the processing and
interpretation of the photometric monitoring data avalanche
from the new facilities allows us to exploit them to the fullest.
However, the nature of a large fraction of the imaging-identified
variables remains unclear unless follow-up spectroscopy is
secured.

\section{Widefield Spectroscopic Telescope}

The evolution from synoptic imaging to synoptic spectroscopy
was prevented by the relatively low multiplexity of spectroscopic
facilities, typically numbering up to a few thousand optical
fibers and by the small field of view of integral field units
(IFUs), usually extending to less than an arcminute. Only
recently a new facility has been considered, WST \citep[Widefield
Spectroscopic
Telescope;][]{2024SPIE13094E..1OB,2024arXiv240305398M} which
increased these parameters by an order of magnitude.

WST is still undergoing definition, but it is expected to be a
10-12-metre telescope, with 3-square-degree field of view,
feeding 20-30 thousand fibers and a 3$\times$3 square-arcmin
IFU. Two resolving power setups are foreseen: a few thousands
and a few tens of thousands. The original plan calls for a
5-year dedicated survey mode operation, delivering hundreds of
millions of spectra with the fibers and a few billion with
the IFU.

The full survey mode operation and the unprecedented level of
multiplexity open up the possibility -- for the first time
and on large scale -- to monitor the spectra of objects for
variability, and to issue spectroscopic alerts, based on the
changes in line parameters: width, flux, velocity shift (and
their ratios for different lines), the appearance and
disappearance of line components, and profile shape.
With some difficulties, continuum parameters can be monitored
too, including shape, slope and even level, provided that a
reasonably accurate aperture-loss control can be achieved.

The list of spectroscopically variable targets is extremely
long and includes a diverse set of objects, from young and
evolved stars (including Be, WR and LBV stars) to a slew of
novae and supernovae, and to changing-look active galaxies.
A comprehensive summary is given in the
project's White Book \citep{2024arXiv240305398M}.

For comparison, the latest ESO multi-object facility, 4MOST
\citep[4-metre Multi-Object Spectroscopic
Telescope;][]{2019Msngr.175....3D} is mounted on a 4-metre
telescope and observes with 1624 fibers feeding a resolving
power 6,500 spectrographs, and with 812 fibers feeding another
spectrograph with a resolving power of 20,000.

A critical requirement for the successful operation of a WST
alert system is to implement an integrated Data Flow System
(DFS) that will process the data on a timely manner and
identify statistically significant changes in the spectra.

\begin{figure*}[p]
\centering
\vspace{-7mm}
\includegraphics[width=17.5cm]{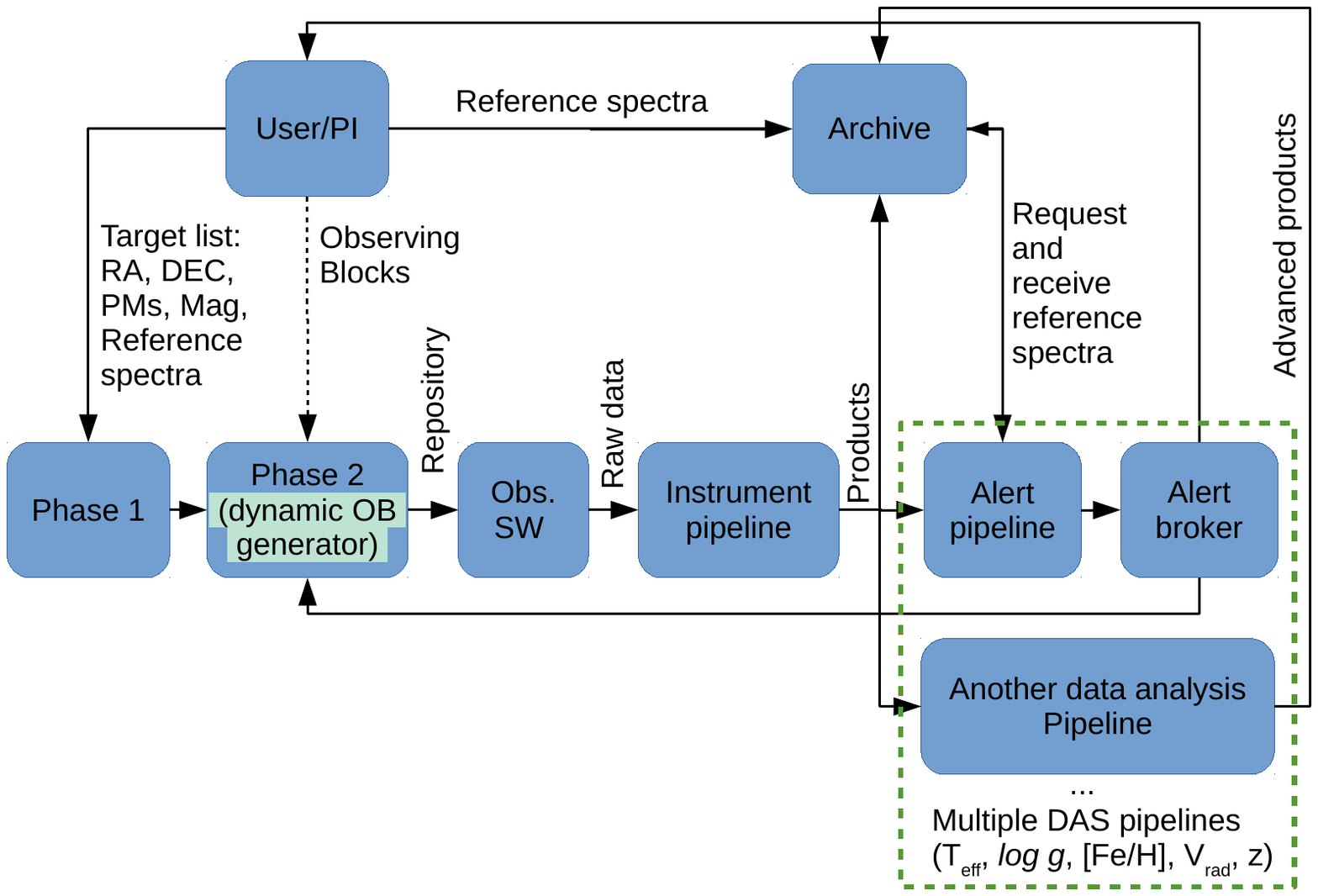}
\vspace{-7mm}
\includegraphics[width=17.5cm]{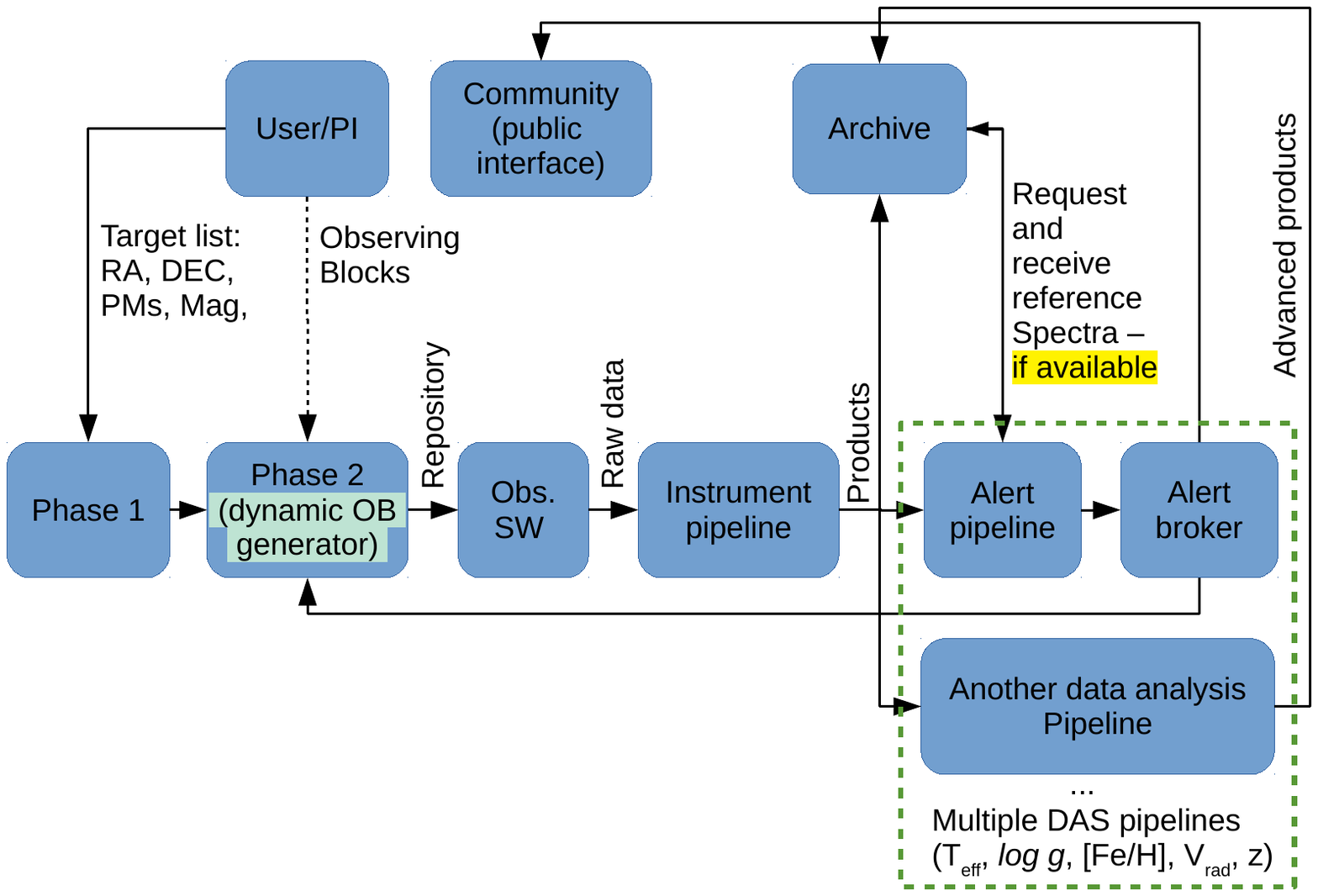}
\caption{Alerts in the WST DFS for different use cases. See
Sec.\,\ref{sec:user_cases} for details.}
\label{fig:user_cases}
\end{figure*}

\section{User Cases and Implications for the Data Flow
System}\label{sec:user_cases}

The WST is envisioned to be a multipurpose spectroscopic
survey facility. Therefore, the alert science should not
disturb any other science goals and this is best achieved
by adopting a modular DFS, separating the basic data reduction
that removes the instrument signature from various high level
analysis pipelines; the variability alert pipeline is one of
the many such pipelines. It will include one or more brokers
which classify, evaluate and rank the identified alerts, and
contact the relevant users.

Two scenarios of variability detection are possible in the
course of the WST operation:
\begin{itemize}

\item The target is any kind of known, expected or suspected
variable source and investigating its spectral variability is
a science goal of the current observation. This is a typical
monitoring program. In the extreme case the study may be
limited to obtaining only one new epoch, or it may consist of
hundreds of epochs.
In this case, as a matter of course, there will be a reference
spectrum to compare the new observations with. The reference
spectra can originate from the WST itself, having been taken
prior to the current observations, or they can be spectra from
other facilities or even a theoretical model (if appropriate),
supplied by the user. This implies that the Phase 1 and 2 tools
can maintain and transfer spectral data or information about
the location of the spectra in the archive, possibly with tools
not unlike the parameter files (so-called PAFs) that are used in
the ESO VLT software to carry various configuration parameters
across the DFS.

\item The target is a random object that has not been flagged
for variability. This will be the case with the vast majority
of observations. However, many of them may have previously
taken archival spectra, even though the spectroscopic variability
is not a science goal of the current observing program. To take
advantage of this data for the purpose of time-domain astronomy
two conditions must be met: (I) the pipeline should be able to
inquire with the WST archive or other Virtual Observatory
repositories, including other telescope archives, whether a target
has previous observations, and (II) the WST archive should ingest
other spectroscopic surveys, at least the major ones, like SDSS
\citep[The Sloan Digital Sky Survey, its Southern
footprint;][]{2000AJ....120.1579Y},
4MOST \citep{2019Msngr.175....3D}, etc. While this may seem
like an overly ambitious goal at present, two decades down the
road the progress in computing and data storage technologies
may render it a trivial task. Furthermore, the future advanced
Virtual Observatory tools may offer a viable alternative.
Last but not least, for any random object, the pipeline can
calculate from the flux-calibrated spectra any synthetic colors
and magnitudes. Conditional upon a reliable aperture-loss
parametrization, this will open the possibility to integrate
the WST with the network of most optical (and possibly near
infrared) imaging surveys, because any filter with a transmission
region that falls within the range of the WST spectral coverage
can be reproduced from the spectra and the atmospheric
transmission curve.

\end{itemize}

From the point of view of the DFS the two cases can
be handled in a uniform manner, if the comparison spectra are
stored in the WST archive. This has the advantage of eliminating
the transfer of the actual data across the system -- only the
location or identification of the reference spectra will have to
be carried across. The submitted reference spectra do not have
to be accessible to the general archive user -- the ESO archive
offers an example, it has a system to control the access to every
piece of data, depending on proprietary periods, for instance.
Restricting access to the PI-submitted reference spectra is
important, because they may not have been published.

The implementation of the integrated alert pipeline and WST
archive is shown in Fig.\,\ref{fig:user_cases}, and indeed, the
two schemes are nearly identical, except for a few minor
differences:

\begin{itemize}

\item In case of variability studies the users must supply a
reference spectrum through a dedicated archive interface, not unlike
the Phase
III\footnote{\url{https://www.eso.org/sci/observing/phase3.html}}
interface that works currently at ESO.

\item The alert pipeline reports the alert to the principal
investigator (PI) of the current program (or to a co-PI to whom
this function has been delegated) in the first case, and to the
wider community in the second.

\item The request for a reference spectrum that the post-processing
alert pipeline sends to the archive is guaranteed to return a
reference spectrum in the first case, but it may return nothing in
the second case.

\end{itemize}

\begin{figure*}[h!]
\centering
\includegraphics[width=17.5cm]{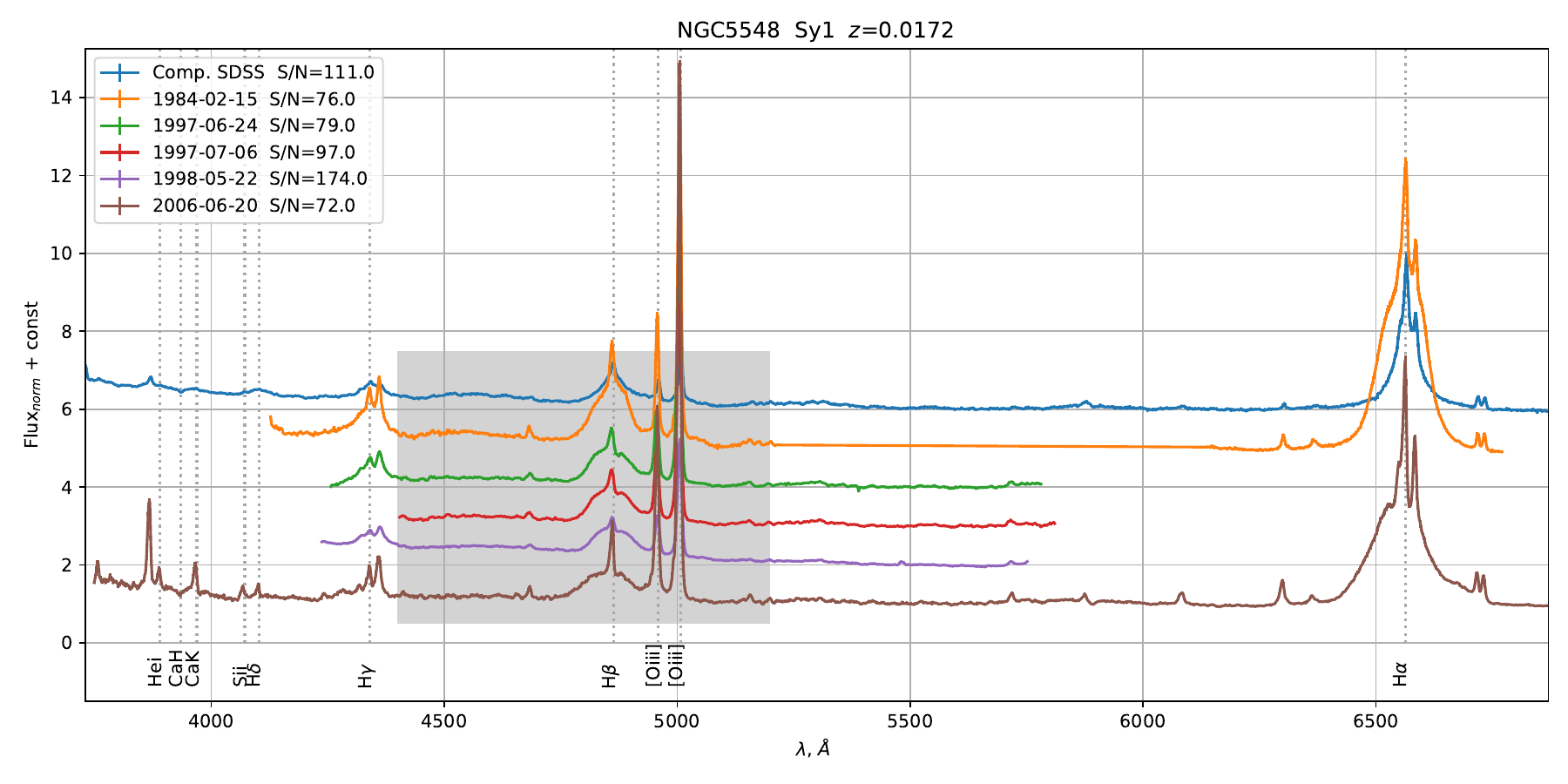}
\includegraphics[width=17.5cm]{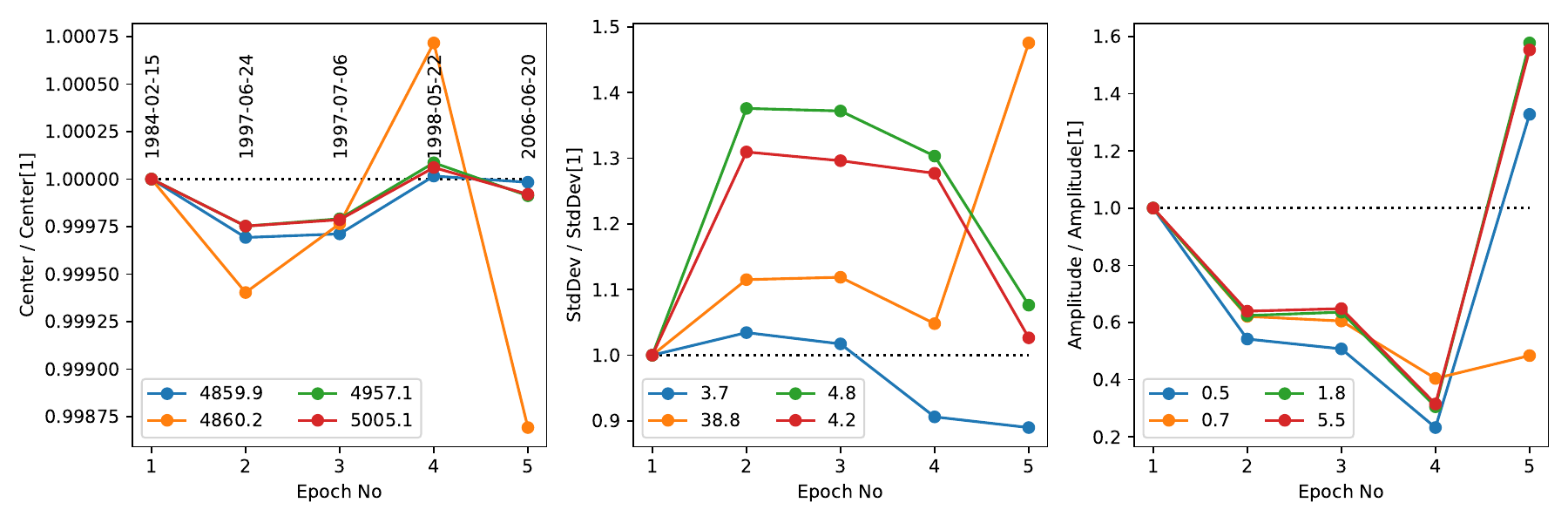}
\caption{Spectral fitting and parameter evolution for the Seyfert 1
galaxy NGC\,5548. See Sec.\,\ref{sec:physics} for details.}
\label{fig:spec_fits}
\end{figure*}

The DFS, the alert pipeline and the associated brokers must work
reasonably fast and the exact requirements depend on the class
of objects. The variable sources that were considered for the WST
can change on time scales from minutes to years. Perhaps, the
most demanding requirement comes from stellar flares that can
have a critical effect on the exoplanet habitability, especially
in the case of late-type stars. These events can be extremely fast,
lasting a few minutes.
Their follow-up may require hierarchical prioritozation in the
data reduction to speed up the analysis of potentially flaring
stars, so it is completed in the course of a single exposure and a
decision can be made by the broker to continue observing with the
same telescope
pointing and configuration until the flare is over -- should such
a program be deemed important enough and ranked high enough to
overwrite the short-term scheduling and effectively issue a
telescope-level ToO (target of opportunity) request.

Finally, seamless integration into the existing ESO
infrastructure -- if the WST indeed is adopted as the next big
project of the observatory -- and Virtual observatory compliance
are needed for faster and easier implementation of the demanding
alert system in the wide landscape of astronomical tools.

\section{Physical Interpretation of the Changing
Spectra -- A Primer}\label{sec:physics}

Here we propose a novel method for detecting changes in the newly
observed spectra. The traditional approach is to compare them with
reference spectra directly, perhaps with some continuum level and
shape adjustments, and radial velocity shifts. However, even
considering the flux errors, the detected changes can only be
evaluated and ranked in one aspect -- their statistical
significance. Ranking the alert will be very important, when the
WST begins to issue hundreds or thousands of alerts every night.

An alternative approach is to model the observed spectra and to
parametrize them in terms of physical units: fit the lines with
appropriate profiles (Gaussian, Lorentz, instrumental, etc. or a
combination of multiple profiles), and measure their fluxes in
energy units, measure the line widths and shifts in km\,s$^{-1}$,
and the continuum level -- also in energy units. An identical
analysis must be performed for both the observed and the reference
spectra, then the comparison can be performed in the space of the
fitting parameters. This comparison will go beyond the statistical
significance, to the physical significance of the changes. For
example, two equal changes in line widths of 100\,km\,s$^{-1}$ and
with equal statistical significances will have different importance
if the lines they were measured on have widths of 2000\,km\,s$^{-1}$
and 12,000\,km\,s$^{-1}$.

An example of such analysis is shown in Fig.\,\ref{fig:spec_fits}
for NGC\,5548, a Seyfert 1 galaxy that is known to show variable
broad emission lines \citep{1999ApJ...524...71K,2000ApJ...536..284K}.
Archival spectra come from
\cite{1995ApJS...98..477H,1999PASP..111..438F,2013ApJ...779..109P,2017cos..rept....4P,2022ApJS..261....4O}
and span a period of over two decades. For this demonstration we
concentrated on the region around 4400-5200\,\AA, encompassing the
two-component H$\beta$ (broad and narrow) and two [Oiii] lines
(top panel). The lines were fitted with Gaussian profiles and the
evolution of each line's center, width and amplitude are also shown
(bottom panel). The differences between the spectra are significant,
but given the heterogeneous nature of the data, we refrain from
drawing further conclusions without a study of the finer
instrumental effects, such as instrumental line profiles and
quality of the wavelength calibration.

Of course, if the reference spectra originate from the WST itself,
these problems will be greatly diminished, but even if the reference
spectra were obtained at another facility, given a large enough suite
of overlapping observations one can hope to parametrize these
differences.

The comparison in physical space will not be applicable to all WST
targets; most likely, it will only work for those with a known and
well-understood nature.
Furthermore, this will probably not be the only
method for detecting spectral changes. Most likely, multiple alert
detection strategies will be necessary, and updates will be required
in the course of the observations, because of the constantly
evolving new methods and tools. Applying different techniques will
increase the reliability, redundancy, and robustness of the alert
identification.

\section{Summary and Conclusions}\label{sec:summary}

We discuss here how the enormous multiplexity of the proposed WST
allows us to shift from triggering photometric alerts based on flux
changes to spectroscopic alerts based on much finer details: line
shapes, widths and strengths (including their ratios), as well as
continuum
level and shape. The challenges in terms of data reduction speed
and DFS complexity are enormous, but so is the potential new
parameter space for discoveries. Finally, we propose an alternative
to the simple statistical comparison of the newly obtained and the
reference spectra: fitting them, thereby
moving the comparative analysis to
the space of physical units, derived from this fit. This will add
context and allow for better alert prioritization -- an important
issue for deciding which alerts must be followed up.

\section*{Acknowledgements}
The author is grateful to all colleagues in the WST project for the
fruitful discussions.

{\small
\bibliographystyle{abbrvnat}
\setlength{\bibsep}{-2pt}
\bibliography{biblio_alerts}
}

\end{document}